\documentclass[aip,amsmath,amssymb,reprint,]{revtex4-1}

\usepackage{graphicx}
\usepackage{amssymb,amsfonts,amsmath}
\usepackage{color}
\usepackage{ulem}
\usepackage[hidelinks]{hyperref}
\usepackage{bbm}

\usepackage{dcolumn}
\usepackage{bm}

\usepackage[utf8]{inputenc}
\usepackage[T1]{fontenc}
\usepackage{mathptmx}
\usepackage{etoolbox}

\usepackage{tikz}
\usetikzlibrary{shapes}
\usetikzlibrary{patterns}
\usetikzlibrary{angles, quotes}
\definecolor{bostonuniversityred}{rgb}{0.8, 0.0, 0.0}
\definecolor{chromeyellow}{rgb}{1.0, 0.65, 0.0}

\newcommand{\rv}{{\mathbf r}}
\renewcommand{\vec}{\mathbf}
\newcommand{\rhotwo}{\rho^{(2)}}

\usepackage{amssymb}
\usepackage{bbold}
\newcommand{\unit}{\vec{e}}

\makeatletter
\def\@email#1#2{%
 \endgroup
 \patchcmd{\titleblock@produce}
  {\frontmatter@RRAPformat}
  {\frontmatter@RRAPformat{\produce@RRAP{*#1\href{mailto:#2}{#2}}}\frontmatter@RRAPformat}
  {}{}
}%
\makeatother

\begin{document}


\preprint{AIP/123-QED}

\title{Superadiabatic dynamical density functional study of Brownian\\ hard-spheres in time-dependent external potentials}

\author{S.~M. Tschopp}
\email{salomee.tschopp@unifr.ch}
\affiliation{Department of Physics, University of Fribourg, CH-1700 Fribourg, Switzerland}
\author{H.~D. Vuijk}
\affiliation{Institute of Physics, University of Augsburg, 86159 Augsburg, Germany}
\author{J.~M. Brader}
\affiliation{Department of Physics, University of Fribourg, CH-1700 Fribourg, Switzerland}

\date{\today}

\begin{abstract}
Superadiabatic dynamical density functional theory 
(superadiabatic-DDFT), a first-principles approach based on the inhomogeneous two-body correlation functions, is employed to investigate the response 
of interacting Brownian particles to time-dependent external 
driving. 
Predictions for the superadiabatic dynamics of the one-body density are made directly from the underlying interparticle interactions, 
without need 
for either adjustable fit parameters or simulation input. 
The external potentials we investigate have been chosen to 
probe distinct aspects of structural relaxation in dense, 
strongly interacting liquid states.
Nonequilibrium density profiles predicted by the 
superadiabatic theory are compared with those obtained from  
both adiabatic DDFT and event-driven Brownian dynamics simulation. 
Our findings show that superadiabatic-DDFT accurately predicts the time-evolution of the one-body density.  
\end{abstract}

\maketitle

\section{Introduction}

Classical Density Functional Theory (DFT) is an
exact framework for determining the 
equilibrium microstructure and thermodynamics of classical
many-particle systems in external fields 
\cite{Evans79,Evans92}. 
The standard method to obtain the one-body density within DFT is to minimize the grand potential functional and solve the resulting Euler-Lagrange (EL) equation for a specified 
external potential at the temperature and chemical potential 
of interest. 
When the grand potential functional is not exactly known, 
as is usually the case, the EL equation yields 
density profiles consistent with the compressibility route 
to the thermodynamics \cite{Evans92,tschopp3}. 
An alternative approach is provided by the force-DFT 
in which the same grand potential functional 
is used to obtain density profiles consistent with the virial route \cite{tschopp3}. 
The key feature of force-DFT is that it explicitly uses the inhomogeneous {\it two-body} density and exploits the fact that this quantity is a functional of the one-body density.
Substitution of the two-body density functional into the 
well-known
%
%
Yvon-Born-Green (YBG) 
equation \cite{Hansen06,mcquarrie} then yields  
a self-consistent scheme for determining the equilibrium 
one-body density. 
Inconsistency between the compressibility and virial routes 
is familiar from integral equation theories 
of bulk fluids \cite{caccamo}. 
Standard DFT and force-DFT present an analogous situation 
on the level of the inhomogeneous one-body density profile, as
discussed in detail in reference \cite{tschopp3}.

The simplest generalization of DFT to 
treat nonequilibrium Brownian systems is known as the 
dynamical density functional theory (DDFT). 
In analogy with equilibrium DFT there are 
two operationally distinct, but formally equivalent, 
variations of DDFT.
While both of these are based on the assumption 
that the nonequilibrium two-body density can be represented 
by the equilibrium two-body density corresponding to the 
instantaneous nonequilibrium one-body density, they differ 
in how this `adiabatic approximation' is implemented. 

Within standard DDFT the dynamics of the one-body density 
are driven by the gradient of the one-body direct correlation 
function \cite{Evans79,Marconi98,ArcherEvansDDFT}, whereas the 
force-DDFT involves explicit integration of the adiabatic 
two-body density to obtain the average interaction force 
at each time-step \cite{tschopp3}. Although these approaches often yield qualitatively reasonable results, they are not quantitatively 
reliable and can break down completely in situations for which 
the time-evolution of the microstructure involves strongly 
correlated particle motion \cite{Marconi98,biglongreview}.

Very recently a first-principles superadiabatic-DDFT 
has been developed and implemented \cite{tschopp4}. 
This new approach yields the 
leading order correction to the adiabatic approximation by explicitly addressing the nonequilibrium dynamics of the 
inhomogeneous two-body density. 
The improved resolution provided by the two-body correlations 
allows for a more realistic description of 
structural relaxation in strongly interacting systems. 
%
%
%
Superadiabatic-DDFT does not involve an explicit memory 
kernel, but rather encodes the history of the system into 
the current values of the one- and two-body 
density. 
The time-evolution of these quantities is determined by the 
simultaneous solution of a pair of coupled, time-local differential equations. 
By identifying the one- and two-body density as relevant variables the superadiabatic-DDFT enables a 
self-contained, autonomous description of many-body Brownian dynamics, which captures the dominant physical processes in 
the system. 
Explicit treatment of the inhomogeneous two-body density provides detailed information about the internal structure of the fluid, from 
which integrated quantities such as the stress tensor 
\citep{Irving_Kirkwood,AerovKrueger} or one-body current can 
be calculated at any point of the time-evolution.  
%
%

In reference \cite{tschopp4} the superadiabatic-DDFT was derived 
by a systematic coarse-graining of the many-body Smoluchowski equation. As a first application, this approach was used to predict the time-evolution of the one-body density profile 
of hard-spheres following an instantaneous change in the 
external potential. 
The agreement between the theoretical density profiles and 
the Brownian dynamics simulation data was very promising. 
In this paper we continue our investigation of the superadiabatic-DDFT by considering systems driven out-of-equilibrium by the application of a periodically varying time-dependent external potential. 
The two situations considered have been carefully 
selected to probe distinct aspects of structural rearrangement in dense fluids. 
Particular attention will be paid to the transient dynamics 
of the one-body density in going from equilibrium to the steady-state. 
The results from superadiabatic-DDFT will be compared with those 
from force-DDFT, which provides the appropriate reference theory for assessing superadiabatic effects, and event-driven Brownian dynamics (BD) simulation, where the simulations were performed following the algorithm proposed in Ref. \cite{scala}.

\section{Theory}\label{sec:Theory}

\subsection{Microscopic dynamics}

For a system of $N$ interacting Brownian particles 
the time-evolution of the configurational probability density, $P(\rv^N\!,t)$, where 
$\rv^N$ represents the set of all coordinates, is given by the Smoluchowski 
equation \cite{dhont}  
\begin{multline}\label{smoluchowski}
\frac{1}{D_0}\frac{\partial P(\rv^N\!,t)}{\partial t}
=
\sum_{i=1}^{N} \nabla_{\vec{r}_i} \cdot \Big( P(\rv^N\!,t)\big(
\nabla_{\vec{r}_i} \ln (P(\rv^N\!,t))  
\\ 
+ \nabla_{\vec{r}_i} \beta U(\rv^N\!,t)
\big) \Big),
\end{multline}
where $\beta\!=\!(k_BT)^{-1}$ and $D_0$ is the bare diffusion coefficient.  
For systems with pairwise interactions 
the total potential energy is given by 
\begin{equation}
U(\rv^N\!,t) = \sum_{i<j}^{N}\phi(r_{ij}) 
+ \sum_{i=1}^{N}V_{\text{ext}}(\rv_i,t),
\end{equation}
where $\phi$ is the pair potential, $r_{ij}=|\vec{r}_i-
\vec{r}_j|$ and 
$V_{\text{ext}}$ is a time-dependent external 
potential.

\subsection{Superadiabatic-DDFT}

The superadiabatic-DDFT, presented in reference \cite{tschopp4}, consists of a pair of differential 
equations for the coupled time-evolution of the one- and two-body densities. 
The first equation is obtained by 
integrating the Smoluchowski equation \eqref{smoluchowski} over $N\!-\!1$ particle coordinates. This yields 
\begin{align} \label{one-body exact}
\frac{1}{D_0}  \frac{\partial \rho(\vec{r}_1,t)}{\partial t} =& \,\nabla_{\vec{r}_1} \!\!\cdot \!\Bigg(\!\nabla_{\vec{r}_1} \rho(\vec{r}_1,t) + \rho(\vec{r}_1,t) \nabla_{\vec{r}_1} \beta V_{\text{ext}}(\vec{r}_1,t) \notag\\
&+ \int d \vec{r}_2 \, \rho^{(2)}(\vec{r}_1,\vec{r}_2,t) \nabla_{\vec{r}_1} \beta \phi(r_{12}) \!\Bigg), 
\end{align}
which is an exact equation of motion for the one-body density, 
$\rho$, requiring as input the nonequilibrium two-body density, 
$\rho^{(2)}$. 

Integrating the Smoluchowski equation \eqref{smoluchowski} over $N\!-\!2$ particle coordinates yields a formally exact equation 
of motion for the two-body density. This includes an integral term involving the nonequilibrium three-body density, which is an unknown quantity. However, by invoking an adiabatic closure the full integral term can be approximated using the 
two-body density \cite{tschopp4}. 
The resulting approximate equation of motion, which constitutes the second equation of superadiabatic-DDFT, is given by
\begin{align} \label{two body adiabatic}
&\frac{1}{D_0} \, \frac{\partial \rho^{(2)}(\vec{r}_1,\vec{r}_2,t)}{\partial t} = \\
&\sum_{i=1,2} \nabla_{\vec{r}_i} 
\!\cdot\! \Bigg( \!\nabla_{\vec{r}_i} \rho^{(2)}_{\text{\,sup}}(\vec{r}_1,\vec{r}_2,t)
\!+\! \rho^{(2)}_{\text{\,sup}}(\vec{r}_1,\vec{r}_2,t) \nabla_{\vec{r}_i} \beta \phi(r_{12}) \notag\\
&\!\!\!+\! \rho^{(2)}(\vec{r}_1,\vec{r}_2,t) \nabla_{\vec{r}_i} \beta V_{\text{ext}}(\vec{r}_i)
\!-\! \rho^{(2)}_{\text{ad}}(\vec{r}_1,\vec{r}_2,t) \nabla_{\vec{r}_i} \beta V_{\text{ad}}(\vec{r}_i,t) \!\!\Bigg) , \notag
\end{align}
where the superadiabatic part of the the two-body density is defined according to
\begin{equation} \label{rho sup}
\rho^{(2)}_{\, \text{sup}}(\vec{r}_1,\vec{r}_2,t) \!\equiv\! \rho^{(2)}(\vec{r}_1,\vec{r}_2,t)-\rho^{(2)}_{\text{ad}}(\vec{r}_1,\vec{r}_2,t) 
\end{equation}
and where the adiabatic two-body density, $\rho^{(2)}_{\text{ad}}$, is 
found by evaluating the equilibrium two-body density at the 
instantaneous one-body density
\begin{equation} \label{rho2 ad}
\rho^{(2)}_{\text{ad}}(\textbf{r}_1,\textbf{r}_2,t)
\equiv
\rho^{(2)}_{\text{eq}}(\textbf{r}_1,\textbf{r}_2;
[\rho(\textbf{r},t)]). 
\end{equation}
Equation \eqref{rho2 ad} expresses a concept essential 
for understanding the coupled superadiabatic-DDFT equations 
\eqref{one-body exact} and \eqref{two body adiabatic}, namely 
that the equilibrium two-body density is a {\it functional} of 
the one-body density \cite{tschopp1,tschopp3,tschopp4}. 
The adiabatic two-body density is obtained by evaluating the 
equilibrium two-body density functional using the 
instantaneous nonequilibrium one-body density. 
The adiabatic potential, $V_{\text{ad}}$, appearing in 
\eqref{two body adiabatic} generates the fictitious external force field required to stabilize the adiabatic system.  
This is given by the YBG relation of equilibrium 
statistical mechanics \cite{Hansen06,mcquarrie},
\begin{multline} \label{V ad} 
-\nabla_{\vec{r}_1} V_{\text{ad}}(\vec{r}_1,t) 
\equiv
k_BT\,\nabla_{\vec{r}_1} \ln \rho(\vec{r}_1,t) 
\\ 
+ \int d \vec{r}_2 \, 
\frac{\rho_{\text{ad}}^{(2)}(\vec{r}_1,\vec{r}_2,t)}
{\rho(\vec{r}_1,t)}
\nabla_{\vec{r}_1}\phi(r_{12}), 
\end{multline}
but here applied to the nonequilibrium system.

For given interaction potential and external field 
the superadiabatic-DDFT predicts the coupled time-evolution 
of the one- and two-body density, starting from their initial values $\rho(\rv_1,t\!=\!0)$ and $\rhotwo(\rv_1,\rv_2,t\!=\!0)$. 
The theory has no adjustable fit parameter and is not dependent 
on any input from stochastic simulations (the BD simulation data shown later in this work are purely for 
comparison). 
Although the superadiabatic theory is not restricted to 
any particular choice of external field we will focus 
in this paper on external potentials exhibiting planar geometry, for which the one-body density varies only as a 
function of a single cartesian coordinate.
%
In this case the equations of superadiabatic-DDFT can be 
simplified, as discussed in detail in Section III.A. of Reference \cite{tschopp4}.

\subsection{Equilibrium limit}\label{subsection equil limit}

In equilibrium the term in parentheses on the right hand-side of equation \eqref{one-body exact} vanishes and we obtain the first-order YBG equation 
\begin{multline} \label{YBG 1}
\nabla_{\vec{r}_1} \rho_{\text{eq}}(\vec{r}_1) + \rho_{\text{eq}}(\vec{r}_1) \nabla_{\vec{r}_1} \beta V_{\text{ext}}(\vec{r}_1)\\ 
+ \int d \vec{r}_2 \, \rho_{\text{eq}}^{(2)}(\vec{r}_1,\vec{r}_2; [\rho_{\text{eq}}]) \nabla_{\vec{r}_1} \beta \phi(r_{12}) = 0. 
\end{multline}
The zero-sum of these three terms expresses the equilibrium balance between 
Brownian, external and internal forces \cite{Hansen06}. 
The notation employed again highlights that the 
equilibrium two-body 
density is a unique functional of the one-body density.  
Of the various methods available to obtain this two-body density functional, the most accurate are 
those based on solution of the inhomogeneous Ornstein-Zernike (OZ) equation
\begin{multline} \label{oz}
h_{\text{eq}}(\rv_1,\rv_2; [\rho_{\text{eq}}]) = c^{(2)}_{\text{eq}}(\rv_1,\rv_2; [\rho_{\text{eq}}]) \\ + \int\! d\rv_3\, h_{\text{eq}}(\rv_1,\rv_3; [\rho_{\text{eq}}])\, \rho_{\text{eq}}(\rv_3) \, c^{(2)}_{\text{eq}}(\rv_3,\rv_2; [\rho_{\text{eq}}]), 
\end{multline}
where $h_{\text{eq}}$ is the total correlation function and 
$c^{(2)}_{\text{eq}}$ is the two-body direct correlation function. 
In this work we choose to follow the approach employed in 
reference \cite{tschopp4} and calculate $c^{(2)}_{\text{eq}}$ 
by taking the second functional derivative of the 
excess (over ideal) Helmholtz free energy functional, $F^{\,\text{exc}}$, with respect to the density 
\begin{equation} \label{c2 functional}
c^{(2)}_{\text{eq}}(\rv_1,\rv_2; [\rho_{\text{eq}}])=-{\frac{\delta^2 \beta F^{\,\text{exc}}[\rho]}{\delta\rho(\rv_1) \delta\rho(\rv_2)}}\bigg\rvert_{\rho_{\text{eq}}}.  
\end{equation}
As there exist reliable approximations to $F^{\,\text{exc}}$ 
for many model systems, we can regard $c^{(2)}_{\text{eq}}$ as a known quantity, such that $h_{\text{eq}}$ can be determined 
by solution of equation \eqref{oz}. 
This can then be related to the two-body density according to 
\begin{equation} \label{rho2 definition}
h_{\text{eq}}(\rv_1, \rv_2; [\rho_{\text{eq}}]) = \frac{\rho^{(2)}_{\text{eq}}(\rv_1, \rv_2; [\rho_{\text{eq}}])}{\rho_{\text{eq}}(\rv_1) \rho_{\text{eq}}(\rv_2)} -1.
\end{equation} 
The closed set of equations \eqref{YBG 1},\eqref{oz},\eqref{c2 functional} and \eqref{rho2 definition} constitutes the force-DFT \cite{tschopp3}, which is the equilibrium limit of superadiabatic-DDFT.

\section{Results for three-dimensional hard-spheres in planar geometry}

The hard-sphere model captures the packing constraints which dominate structural relaxation in dense liquids and thus presents an appropriate choice for the present study. 
In the following we show numerical results for a three-dimensional system of hard-spheres, of diameter $d\!=\!1$, subject to a time-dependent external potential. The units of energy and time are fixed by  $k_BT\!=\!1$ and $d^2/D_0\!=\!1$, respectively.
The external potential considered is a function of a single cartesian coordinate, taken here as the $z$-direction, which 
imposes a planar symmetry.  
The potential consists of a harmonic trap with an additional  
(time-dependent) Gaussian peak and takes the following form
\begin{equation}\label{disco_fox_potential}
V_{\text{ext}}(z,t) 
= A\,\Big(z-z_0^A\Big)^{2} 
+ 
B(t)\,e^{-\alpha \left(z-z_0^B(t) \right)^2},
\end{equation}
where the harmonic trap amplitude is set equal to $A\!=\!1.5$ with its minimum located 
at $z_0^A\!=\!5$. 
The Gaussian decay parameter is set to a constant value of 
$\alpha\!=\!5$, while its amplitude and peak position are 
given by the time-dependent functions $B(t)$ and $z_0^B(t)$, respectively.  
In the following we explore two specific realizations of this 
general external potential to probe strongly correlated cooperative motion and structural relaxation. 
The resulting
one-body density profiles from superadiabatic-DDFT are then 
benchmarked against BD simulation data and 
compared with the predictions of force-DDFT. 
We calculate the adiabatic two-body density 
by using equations \eqref{oz}, \eqref{c2 functional} and 
\eqref{rho2 definition} together with the well-known Rosenfeld 
approximation to the excess Helmholtz free energy 
\cite{rosenfeld89} (see Appendix \ref{rosenfeld_appendix} for 
further details). 

As already mentioned in the introduction, the force-DDFT is a 
purely adiabatic approach which assumes instantaneous equilibration of the two-body density.  
Superadiabatic- and force-DDFT have the same equilibrium 
limit, namely force-DFT, and this convenient feature enables a comparison 
focused solely on adiabatic/superadiabatic differences,  without any residual equilibrium bias. 
A full account of the force-DDFT method can be found in 
Reference \cite{tschopp3}.
Implementation details for both superadiabatic- and force-DDFT 
in planar geometry can be found in Sections III.A. and III.B. of Reference \cite{tschopp4} and in Sections III.G. and IV.A. 
of Reference \cite{tschopp3}, respectively.

\subsection{Periodic compression}

\begin{figure*}
\center{\includegraphics[width=0.9\linewidth]{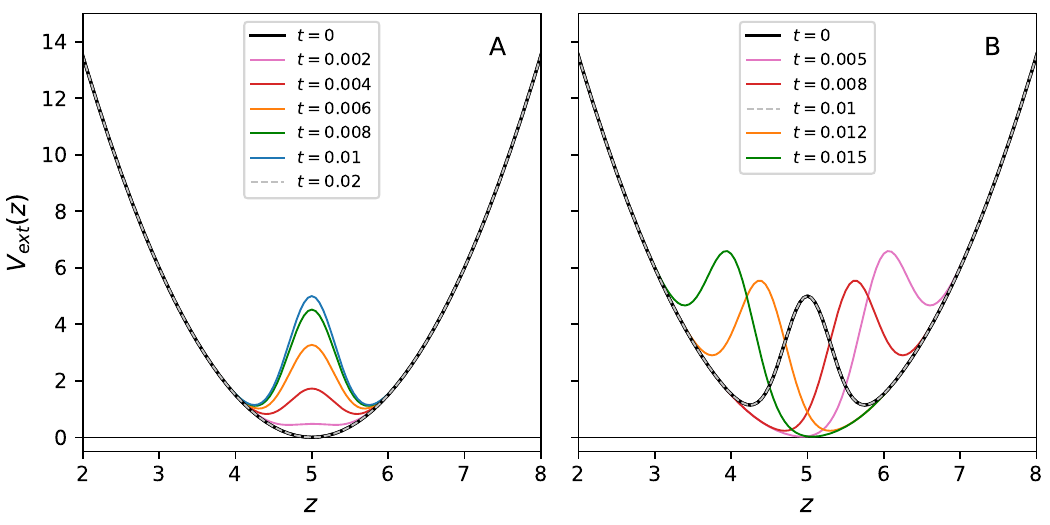}}
\caption{\textbf{Time-dependent external potentials} for 
the two special cases considered in this work. The general form of the potential is given in equation \eqref{disco_fox_potential}. 
In panel A, we show a potential consisting of a confining harmonic trap and a centered Gaussian peak with a time-dependent 
amplitude (see equation \eqref{equation case 1}). 
In panel B, we show a potential consisting of the same harmonic trap, but now with a Gaussian peak which moves (with fixed amplitude) back-and-forth about the trap minimum (see equation \eqref{equation case 2}). 
The first case is designed to `periodically compress' the particles away from the center and towards the outer regions of the trap. The second case induces more flow in the system, as the 
Gaussian peak constantly disrupts the particle microstructure. 
}
\label{figure Vext}
\end{figure*}

For our first test-case we set the external potential \eqref{disco_fox_potential} such that the location of the Gaussian peak is held constant in the center of the harmonic trap, $z_0^B(t)\!=\!z_0^A\!=\!5$, with a time-dependent amplitude given by
\begin{equation}\label{equation case 1}
B(t) = \frac{B_{\text{max}} }{2}\Big( 1 - \cos(\omega t) \Big),
\end{equation}
for $t\!>\!0$ and zero for earlier times.
We choose a maximal amplitude $B_{\text{max}}\!=\!5$ and frequency $\omega\!=\!2 \pi/0.02$. 
The time-dependent variation of the external potential is 
illustrated in panel A of Figure~\ref{figure Vext}.

\begin{figure*}
\center{\includegraphics[width=0.9\linewidth]{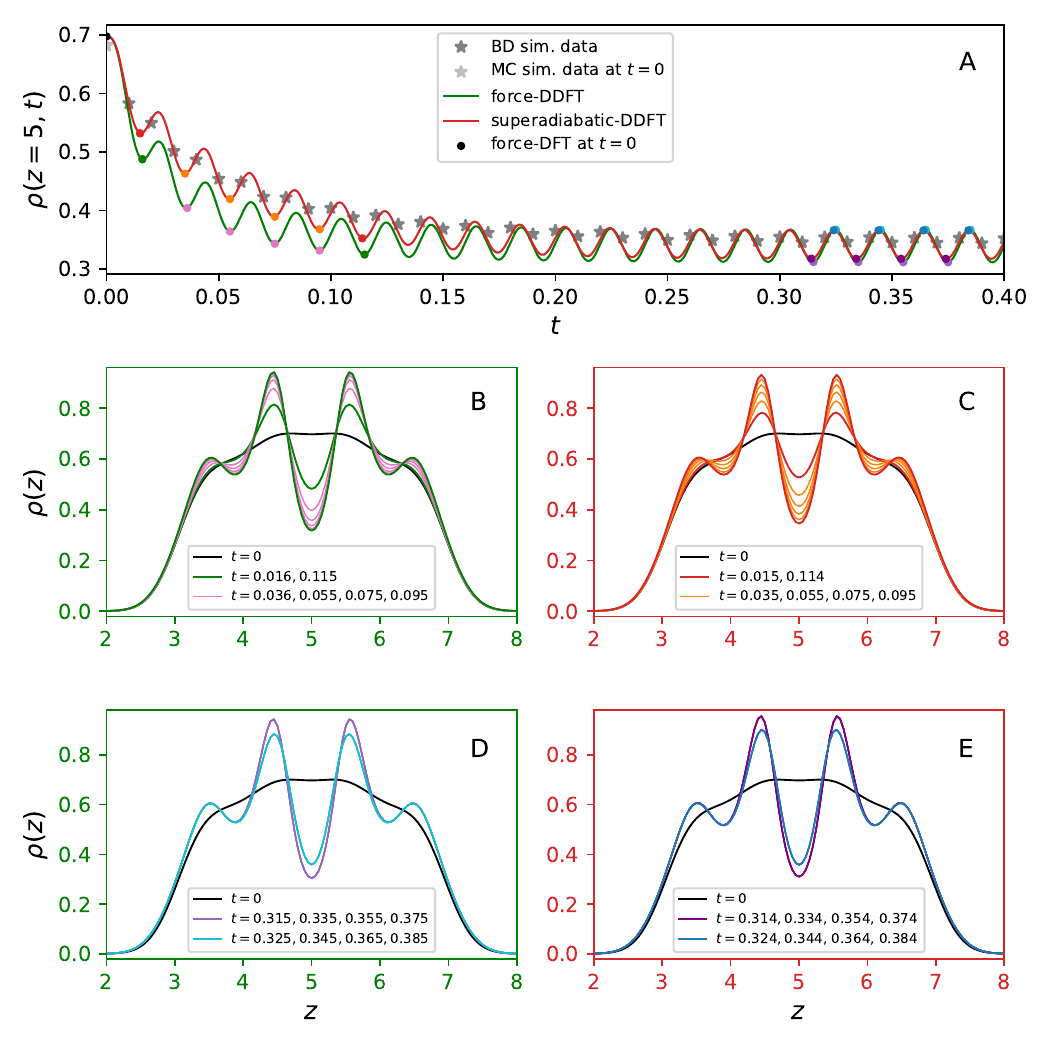}}
\caption{
\textbf{Periodic compression.} Starting from the equilibrium 
density profile in a harmonic trap we 
track the time-evolution of the one-body density following 
the switch-on (at $t\!=\!0$) of an additional repulsive Gaussian potential 
with a time-dependent amplitude, see equations 
\eqref{disco_fox_potential} and \eqref{equation case 1}. 
Panel A shows the density at the trap center 
($z\!=\!5$) as a function of time obtained from 
superadiabatic-DDFT (red line) and force-DDFT (green line). 
The stars are data points from simulation and the filled circles 
indicate the times at which we show full density profiles 
in Panels B to E (note the corresponding colour scheme). 
Following a period of transient relaxation the system 
enters a periodic stationary state for times greater than 
$t\!\sim\!0.2$. 
} 
\label{figure test case 1}
\end{figure*}

The system is initially prepared in an equilibrium state with an 
average of $2.5$ particles per 
unit area in the $xy$-plane, i.e. 
$\langle N\rangle\!=\!\int dz\, \rho_{\rm eq}(z)\!=\!2.5$, before being driven out-of-equilibrium by the oscillatory Gaussian peak \eqref{equation case 1} for 
times $t\!>\!0$.  
Panels B to E of Figure~\ref{figure test case 1} show theoretical 
density profiles at different times, calculated using both superadiabatic-DDFT (red-framed panels C and E) and force-DDFT (green-framed panels B and D). 
The data shown in Panel A tracks the value of the one-body density at the 
center of the harmonic trap, $\rho(z\!=\!5,t)$, as a function of time. 
This shows clearly the transient relaxation of the one-body density from equilibrium to a periodic nonequilibrium steady-state. 
The filled circles indicate times at which we show the full density profile in panels B to E and the stars are results of BD simulation, shown at regular time intervals.   

We first discuss the data shown in Panel A of Figure~\ref{figure test case 1}.
At time $t\!=\!0$ the system is in equilibrium and both 
superadiabatic- and force-DDFT predict identical density 
profiles, namely those of force-DFT. 
The equilibrium density profile is not in perfect agreement with 
that of simulation (calculated using the Monte-Carlo method) due to the approximate free energy used to 
calculate the equilibrium two-body density, in analogy with the 
situation considered in reference \cite{tschopp4}.  
From $t\!=\!0$ until around $t\!=\!0.2$ we observe a transient relaxation during which $\rho(z\!=\!5,t)$ decreases after each oscillation period. 
This decrease reflects the changes in the 
microstructure caused by periodically compressing the 
particles away from the center and towards the outer regions of the harmonic trap.
During the transient regime, interparticle collisions cause 
the particles to adjust their positions  
in such a way that they can move as freely as possible 
back-and-forth in response to the externally applied forces. 
The time-dependent Gaussian respects the symmetry of the 
trap (mirror symmetric about $z\!=\!5$) and does not induce 
a large amount of flow,
but 
rather leads to more subtle configurational changes as each particle modifies its \textit{local} environment through repeated interaction with its neighbours. 
The absence of strong flow in the system suggests that superadiabatic effects can be expected to be modest. The superadiabatic- and force-DDFT predictions for 
$\rho(z\!=\!5,t)$ are given by the red and green curves, respectively.  
The prediction of superadiabatic-DDFT is in excellent quantitative agreement with the BD data and captures almost 
perfectly the transient behaviour, whereas the force-DDFT 
decays too rapidly. 
The superadiabatic-DDFT correctly implements the 
zero flux condition on the two-body density at 
all times and thus provides a much more realistic account of the 
microstructural rearrangements induced by external forces.   

In Panels B and C we show the full one-body 
density profiles from force- and superadiabatic-DDFT, respectively, at selected times during the transient relaxation. 
These times (indicated by filled circles in Panel A) have been selected such that force- and superadiabatic-DDFT profiles can be compared at equivalent points in the oscillation cycle, 
rather than at strictly equal times. 
As the repulsive Gaussian peak grows in magnitude  
(see Figure~\ref{figure Vext}), particles are forced away 
from the center of the harmonic trap and a first packing peak 
develops on either side of the trap minimum. 
Although qualitatively similar, this process occurs more slowly 
in superadiabatic-DDFT than in force-DDFT, which we attribute to the improved treatment of structural relaxation in the 
superadiabatic theory, as discussed above. 
As the system approaches a steady-state, the density profile develops a second packing peak on either side of the trap minimum. 
Here we again observe that the build-up of packing structure 
takes longer in superadiabatic-DDFT than in force-DDFT.

\begin{figure*}
\center{\includegraphics[width=0.9\linewidth]{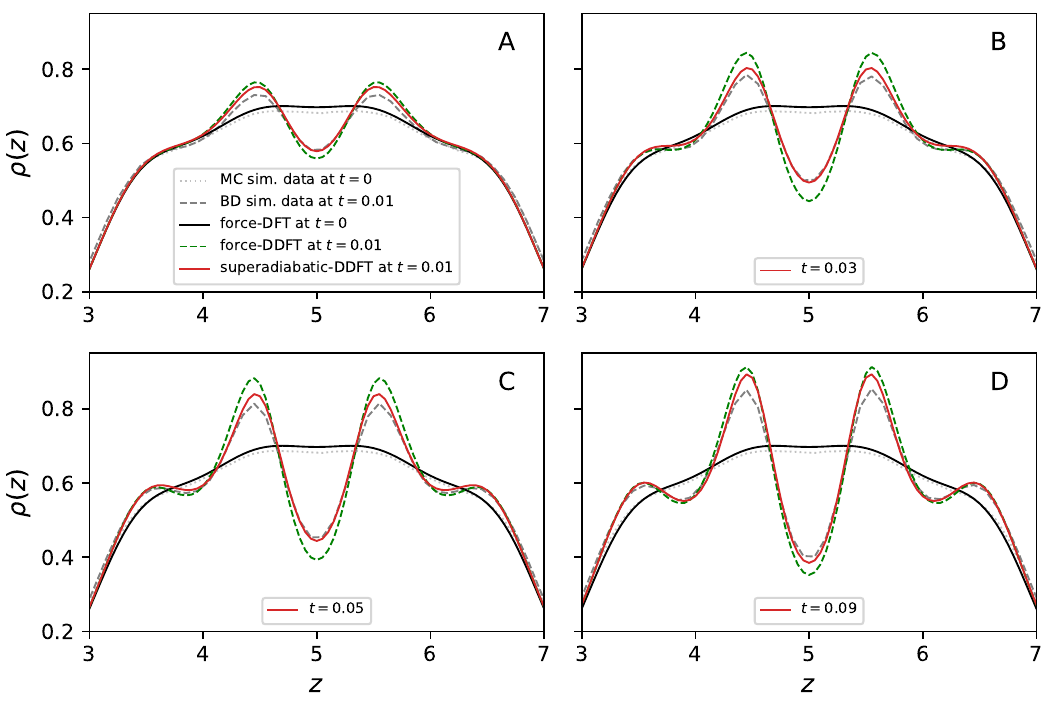}}
\caption{
\textbf{Periodic compression.} Comparing the predictions of force- and 
superadiabatic-DDFT (green dashed lines and red solid lines, respectively) with BD simulation data (gray dashed lines) at four different 
times during the transient time-evolution of the system. 
The external potential is given by 
\eqref{disco_fox_potential} and \eqref{equation case 1} and generates a periodic compression of the particles towards 
the sides of the harmonic trap.
The equilibrium density curves, at $t\!=\!0$, obtained by force-DFT (black solid lines) and Monte-Carlo simulation (silver dottd lines) are included as a reference.
In all cases the superadiabatic-DDFT agrees considerably better 
with the simulation data than the force-DDFT. 
} 
\label{figure test case 1 simulation}
\end{figure*}

Returning to panel A, the value of the one-body density at 
the trap minimum, $\rho(z\!=\!5,t)$, indicates the onset of a steady-state for times $t\!>\!0.2$.   
In Panel E we show full density profiles calculated in this steady-state using superadiabatic-DDFT 
at times separated by one oscillation period. 
The purple coloured profiles were calculated at 
minima of the curve shown in Panel A, whereas the blue profiles were calculated at the maxima. 
The fact that curves of the same colour cannot be 
distinguished from each other confirms that we have indeed  entered a steady-state for the \textit{full} density profile 
and not only for its value at $z\!=\!5$.  
The same conclusion can be drawn from the profiles calculated 
using force-DDFT shown in Panel D. 
The strong similarity between the steady-state profiles calculated using both force- and superadiabatic-DDFT (compare, for example, the 
light and dark purple curves in Panels D and E, respectively) 
is a consequence of a nonequilibrium microstructure which 
allows each particle to oscillate locally back-and-forth without 
very frequent interaction with its neighbours. 
In such a state, one can expect that superadiabatic effects, 
which arise from interparticle interactions, will remain 
small. 
It is thus not surprising that the steady-state density 
profiles of force- and superadiabatic-DDFT are in close agreement. 
This does not apply to the transient regime, during which frequent interparticle interactions serve to 
break-up and rearrange the initial equilibrium microstructure. 
We note that superadiabatic effects will have an increased  influence on the steady-state density profiles in more densely packed systems with a larger value of $\langle N\rangle$, but we choose to focus here on fluid states at more moderate packing.   

\begin{figure*}
\center{\includegraphics[width=0.9\linewidth]{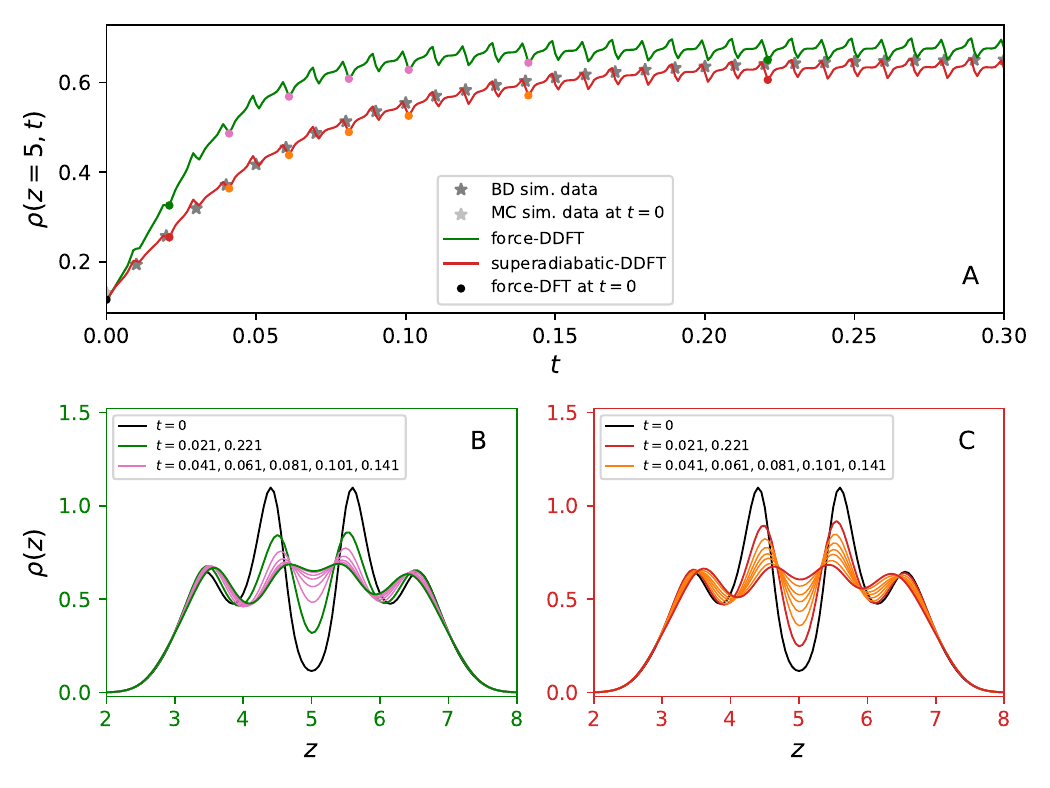}}
\caption{
\textbf{Flow induced mixing.} 
Starting, at $t\!=\!0$, with the equilibrium 
density profile in a harmonic trap with a fixed Gaussian peak 
at $z_0^B\!=\!z_0^A\!=\!5$, we show, for $t\!\ge\!0$, the time-evolution of the one-body density as the Gaussian peak sweeps from side-to-side following equations
\eqref{disco_fox_potential} and \eqref{equation case 2}. 
Panel A shows values of the density at the trap center 
($z\!=\!5$) as a function of time obtained from 
superadiabatic-DDFT (red line) and force-DDFT (green line). 
The stars are data points from simulation and the filled circles 
indicate the times at which we show full density profiles 
within the transient regime in Panels B (force-DDFT) 
and C (superadiabatic-DDFT) - note the corresponding colour scheme.
} 
\label{figure test case 2}
\end{figure*}

In Figure~\ref{figure test case 1 simulation} we compare the 
predictions of force- and superadiabatic-DDFT with simulation 
data at four different times during the transient regime of the 
time evolution. 
The superadiabatic-DDFT is globally more accurate than force-DDFT at all times. 
However, agreement with the simulation data is still not perfect. 
We attribute much of the error exhibited by 
superadiabatic-DDFT to the underlying equilibrium free energy functional used to generate the adiabatic two-body density 
(see equation \eqref{c2 functional}). 
The deviation of the equilibrium force-DFT curve at $t\!=\!0$ from the Monte-Carlo data persists in the nonequilibrium 
density profiles from superadiabatic-DDFT, which is consistent with the findings of reference \cite{tschopp4}. 
We are confident that employing a more accurate equilibrium approximation to generate the adiabatic two-body 
density would enable the superadiabatic-DDFT to give an even 
more satisfactory account of the simulation data. 
In any case, this residual 
equilibrium error becomes less significant when the system undergoes stronger flow and superadiabatic effects become more 
prominent, as demonstrated below.

\subsection{Flow induced mixing}


We next set the external potential \eqref{disco_fox_potential} such that the Gaussian has a constant amplitude, $B(t)\!=\!B_{\text{max}}\!=\!5$, but a 
time-dependent position, given by
\begin{equation}\label{equation case 2}
z_0^B(t) = 
\begin{cases}
z_0^A, & \text{if $t<0$}
\\
z_0^A + \Delta z \sin(\omega t) , & \text{if $t\ge0$}
\end{cases}
\end{equation}
where $\Delta z\!=\!1$ is the maximal displacement of the 
Gaussian peak away from the harmonic potential minimum, located at 
$z_0^A\!=\!5$, and where we have again taken the frequency $\omega\!=\!2 \pi/0.02$. The time-dependent variation of 
this external potential is shown in panel B of Figure \ref{figure Vext}.
The system is initially prepared in an equilibrium state with an average of 2.5 particles per unit area in the $xy$-plane 
and is then driven out-of-equilibrium by the side-to-side oscillatory motion of the Gaussian peak. 

We begin by discussing Panel A of Figure \ref{figure test case 2}, which shows the value of the one-body density at the center of the harmonic trap, $\rho(z\!=\!5,t)$, as a function of time. 
The green and red curves show the predictions of force- and 
superadiabatic-DDFT, respectively, where both theories 
predict a regime of transient relaxation before arriving 
at a nonequilibrium steady-state.
The stars are data from simulation, sampled at regularly spaced time intervals, and the filled circles indicate the times at which we show full density profiles in Panels B and C. 
The predictions of force- and 
superadiabatic-DDFT shown in Panel A reveal much greater discrepancy between the two theories than in the previously considered test-case (see Figure \ref{figure test case 1}).
The superadiabatic-DDFT is once again in very good agreement 
with the BD simulation data and gives an accurate 
description of both the transient relaxation and the steady-state. 
In contrast, the transient predicted by force-DDFT decays 
too rapidly and stabilizes to an erroneous steady-state.

\begin{figure*}
\center{\includegraphics[width=0.9\linewidth]{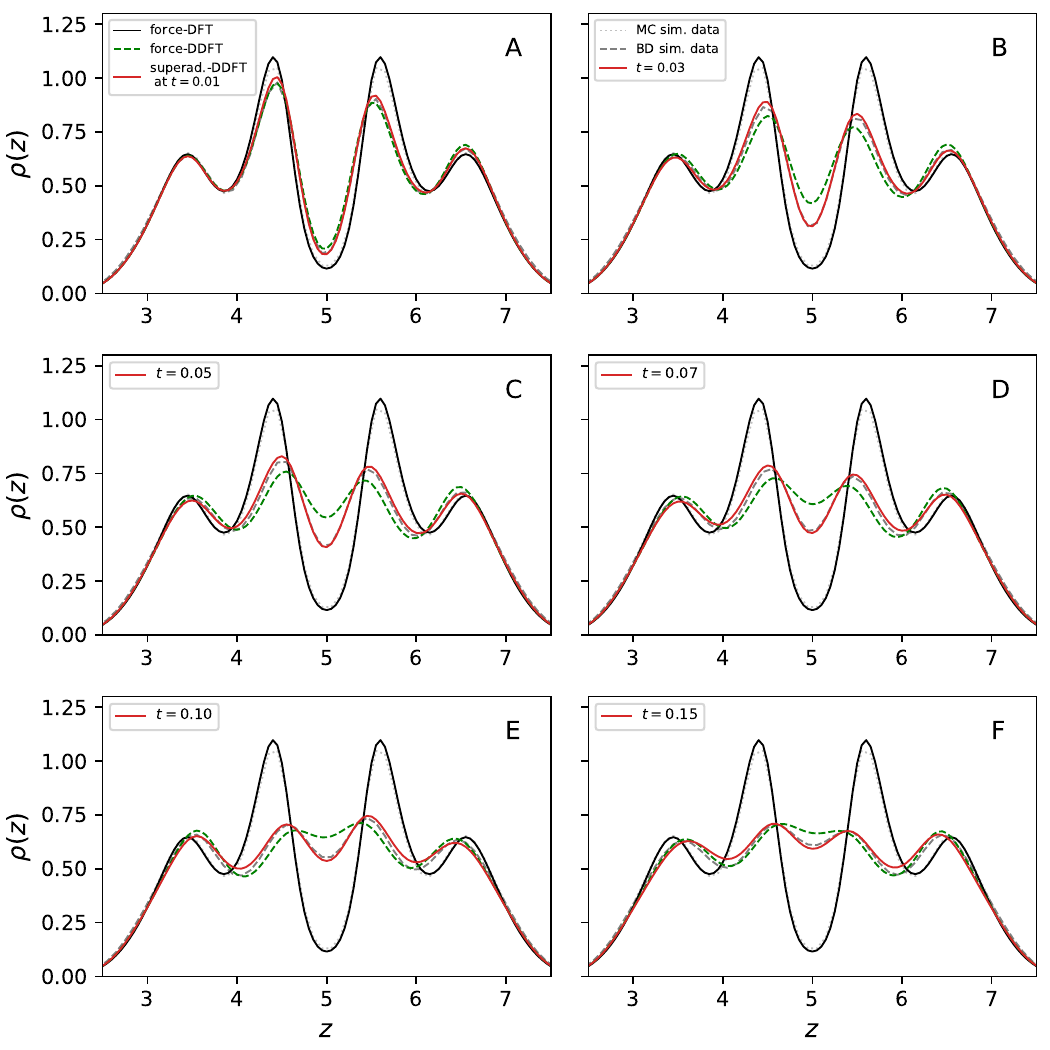}}
\caption{
\textbf{Flow induced mixing.} 
Panels A to F compare theoretical predictions with BD simulation data at different times during 
the transient regime. In each Panel we give equilibrium 
profiles from force-DFT, black solid lines, and Monte-Carlo simulation, silver dotted lines, to serve as a reference. 
Force-DDFT profiles are given by green dashed lines, superadiabatic-DDFT profiles by red solid lines and BD simulation data by dashed gray lines. 
}
\label{figure test case 2 simulation} 
\end{figure*}


While force-DDFT performs rather poorly, giving at 
best a qualitative description, 
the superadiabatic-DDFT gives an excellent quantitative account of the BD data and captures very well the time-scale of transient relaxation. 
The sweeping, back-and-forth motion of the Gaussian peak 
induces a global flow, involving frequent interparticle collisions, which mixes 
the particles as they repeatedly flow over the moving 
potential barrier.   
This continual, global disturbance of the microstructure 
causes the nonequilibrium two-body density to deviate significantly from the adiabatic two-body density and thus 
generates strong superadiabatic effects. 
While the differences between the two theories are most pronounced during the transient, they are still apparent in the steady-state, where the particles continue to undergo numerous collisions as they are forced to move back-and-forth past their neighbours. 
This situation should be contrasted with the steady-state 
of the previous test-case, where superadiabatic effects were found to be small (see Figure \ref{figure test case 1}) due to the relatively unhindered, small-amplitude, oscillatory motion of each particle within its own local environment.

Panels B and C of Figure~\ref{figure test case 2} show theoretical density profiles at different times, calculated using both force-DDFT (green-framed panel B) and 
superadiabatic-DDFT (red-framed panel C). 
The times chosen are indicated by the filled circles in 
Panel A, where we note the matching colour scheme between 
the points and the full density profiles. 
Due to the oscillatory motion of the Gaussian peak, which initially moves to the right following 
the onset of motion at $t\!=\!0$, the density profiles are 
asymmetric about the minimum of the Gaussian trap. 
Consistent with our previous findings the superadiabatic-DDFT reacts slower to changes in the external field than the force-DDFT, e.g.~the central minimum in the equilibrium 
profile is more slowly eroded by the motion of the 
Gaussian peak. 
Indeed, for the later times shown the force-DDFT profiles 
are almost overlapping, reflecting the onset of the steady-state, whereas the corresponding superadiabatic-DDFT profiles 
can be easily distinguished from each other.

In Figure~\ref{figure test case 2 simulation} we compare 
density profiles predicted by superadiabatic- and force-DDFT with BD simulation data. 
As a reference for the eye, we also show the starting equilibrium curves from force-DFT and Monte-Carlo simulation.
In this figure, it is clear that the nonequilibrium density profiles predicted by superadiabatic-DDFT are in remarkably good agreement with the simulation data and capture with high accuracy the 
global shape of the profiles at all times considered. 
In contrast, force-DDFT is not able to 
describe the time-evolution of the profiles in a satisfactory way and yields a generally poor account of the packing oscillations. 

In this second test-case, the improvement of superadiabatic-DDFT over force-DDFT is much more dramatic than in the previous test-case, since the superadiabatic forces, closely linked to the particle packing, stay relevant even into the steady-state regime. For such dynamics explicitly including the time-evolution of the two-body density is very important to describe accurately the microstructural changes in 
the driven system.

\section{Discussion and conclusions} \label{discussion}

In this paper the recently developed superadiabatic-DDFT 
\cite{tschopp4} was used to study the dynamics of a system of hard-spheres subject to a time-dependent external field. 
Our results  
show that superadiabatic-DDFT provides an accurate 
description of structural relaxation in fluids and 
reveal clearly the deficiencies of the more simplistic 
force-DDFT method \cite{tschopp3}.

Within the force-DDFT the adiabatic two-body density is calculated at each time-step and used to generate the one-body 
force due to interparticle interactions. 
However, an undesirable consequence of applying the adiabatic approximation at the one-body level is that both $\rho$ and 
$\rho^{(2)}_{\text{ad}}$ follow an unphysical time-evolution, which does not respect 
the packing constraints imposed by interaction potentials with a strong short-range repulsion.  
More concretely, if we consider the time-evolution of the full many-particle system, as calculated in BD simulation, then 
the functions $\rho$ and $\rho^{(2)}_{\text{ad}}$ predicted 
by force-DDFT could not be obtained from statistical averages 
over a physically realistic sequence of configurations. 
Even if it were possible to reproduce these functions by back-engineering some artifical 
configuration sequence, then this would likely involve unphysical situations for which the particles overlap. 
In contrast, the superadiabatic-DDFT prevents such unphysical 
overlaps, due to the explicit (non-integrated) appearance of 
the pair interaction potential in equation \eqref{two body adiabatic}. The error in superadiabatic-DDFT 
lies in the neglect of subtle differences between higher-order 
correlations of the adiabatic and nonequilibrium systems.  

An alternative framework for treating superadiabatic 
dynamics is provided by the power functional theory (PFT) 
\cite{PFToriginal,SchmidtYellow}, which formally generalizes the variational approach of DFT to nonequilibrium.  
Within PFT all superadiabatic effects are described 
by an excess power functional with a nonlocal dependence 
on the entire history of both the one-body density and current. 
In principle, knowledge of this excess functional would 
provide a closed, predictive theory for the dynamics of 
the one-body density. 
However, in practice, the complexity of the excess functional 
has prevented the construction of any approximation 
capable of providing this closure. 
The fact that PFT requires a temporally nonlocal excess power
functional is a natural consequence of using the one-body density and current as relevant variables. 
The information lost in coarse-graining to these one-body 
fields is compensated by the use of memory kernels,
which are generally difficult to approximate.   
The superadiabatic-DDFT approach employed in this work 
identifies the one- and two-body densities as the relevant 
variables for a coarse-grained description of many-body Brownian 
dynamics. 
This provides detailed information about the microstructure 
of the nonequilibrium fluid and enables the formulation of a closed, time-local dynamical theory.
Within superadiabatic-DDFT the flow history of the system 
is encoded in the current value of the nonequilibrium two-body density, without the need for a memory kernel.

Since collective motion in dense fluids, whether in- or out-of-equilibrium, is dominated by the repulsive part of the interparticle interaction potential, we chose in the present work to focus on the three-dimensional hard-sphere model. However, we emphasise that our approach is by no means limited to hard-spheres and can be applied without difficulty to systems 
interacting via any pairwise additive interaction potential.


Two different time-dependent external potentials were considered. 
In the first case the external field generated a local, periodic compression of the system for which superadiabatic effects are expected to be modest, such that the steady-states predicted 
by superadiabatic- and force-DDFT are very similar.  
This enabled us to focus solely on the transient regime and demonstrate the improved performance of the superadiabtic-DDFT over force-DDFT. 
In the second case the external field
induced a global flow in the system, which generates strong superadiabatic effects in both the transient regime and the steady-state. 
We again found that the superadiabatic-DDFT predicts one-body density profiles in very good agreement with the BD simulation 
data and performs far better than the force-DDFT.

We thus conclude that superadiabatic-DDFT provides a reliable 
and accurate method to predict the dynamics of the one-body density in systems driven by time-dependent external potentials.

\acknowledgments

We thank G.T. Hamsler for general support and deep insights.
We wish her a good retirement.

\section*{Data Availability}

The data that support the findings of
this study are available from the
corresponding author upon reasonable
request.

\appendix

\section{FMT for the two-body density}\label{rosenfeld_appendix}

Within the fundamental measure theory (FMT) the excess Helmholtz free energy functional is approximated by an integral over a function of weighted densities \cite{rosenfeld89}
\begin{align}\label{chap1_ros_fe}
\beta F^{\text{exc}}_{\text{hs}}[\,\rho\,] = \int \!d\rv_1 \; \Phi \left( \left\lbrace n_{\alpha}(\rv_1) \right\rbrace  \right). 
\end{align}
The original Rosenfeld version of FMT employs the following 
approximate form for the reduced excess free energy density of 
the hard-sphere system
\begin{equation}\label{chap1_Phi_ros_fe}
\Phi = - n_0 \ln(1-n_3) + \frac{n_1 n_2 - \vec{n}_1 \cdot \vec{n}_2}{1-n_3} + \frac{n_2^3 
- 3 n_2 \vec{n}_2 \cdot \vec{n}_2}{24 \pi (1-n_3)^2}.
\end{equation}
The weighted densities are generated by convolution 
\begin{equation}\label{chap1_n_alpha_ros_fe}
n_{\alpha}(\rv_1) = \int \!d\rv_2 \; \rho(\rv_2)\, \omega_{\alpha}(\rv_{12}), 
\end{equation}
where $\rv_{12}\!=\!\rv_1\!-\!\rv_2$ and  
the weight functions, $\omega_{\alpha}$, are given by four scalar functions 
\begin{align}
\omega_3(\rv)&=\Theta(R-|\rv|), \hspace*{0.5cm}
\omega_2(\rv)=\delta(R-|\rv|), \notag\\
\omega_1(\rv)&=\frac{\delta(R-|\rv|)}{4\pi R}, \hspace*{0.54cm}
\omega_0(\rv)=\frac{\delta(R-|\rv|)}{4\pi R^2}, \notag
\end{align}
and two vectors
\begin{align}
\omega_{\mathbf{2}}(\rv)&=\unit_{\rv}\,\delta(R-|\rv|),\hspace*{0.55cm}
\omega_{\mathbf{1}}(\rv)=\unit_{\rv}\frac{\delta(R-|\rv|)}{4\pi R},
\notag
\end{align}
where $\unit_{\rv}\!=\!\rv/|\rv|$ is a unit vector. 
The symbol $\omega$ is used here for all weight functions, with vector functions distinguished from scalar functions by using a bold font index, in accordance with the notation introduced in 
Ref. \cite{tschopp2}. 

Within DFT the equilibrium two-body direct correlation function is obtained 
by taking two functional derivatives of the excess Helmholtz 
free energy, as given in equation \eqref{c2 functional}.
When applied to the approximate Rosenfeld expression 
\eqref{chap1_ros_fe} this yields \cite{tschopp2} 
\begin{equation}
c^{(2)}_{\text{eq}}(\rv_1, \rv_2;[\,\rho_{\text{eq}}\,])
=-\!\sum_{\alpha\beta}\int \!d\rv_3\, 
\omega_{\alpha}(\rv_{31})\,
\Phi^{''}_{\alpha \beta}(\rv_3)\,
\omega_{\beta}(\rv_{32}),
\label{chap1_c_ros_fe}
\end{equation}
where $\Phi^{''}_{\alpha \beta}\!=\!\partial^2 \Phi/\partial n_{\alpha} \partial n_{\beta}$ and the sums over $\alpha$ and 
$\beta$ run over all scalar and vector indices. 
From equations \eqref{chap1_Phi_ros_fe}, \eqref{chap1_n_alpha_ros_fe} and \eqref{chap1_c_ros_fe} it is 
clear that the two-body direct correlation function is determined purely by the one-body density. 
Substitution of a nonequilibrium one-body density into the 
expression \eqref{chap1_c_ros_fe} generates the adiabatic 
two-body direct correlation function 
\begin{equation}
c^{(2)}_{\text{ad}}(\rv_1, \rv_2, t) \equiv
c^{(2)}_{\text{eq}}(\rv_1, \rv_2; [\rho(\textbf{r},t)]).
\end{equation}
Substitution of $c^{(2)}_{\text{ad}}$ into the inhomogeneous 
OZ equation
\begin{multline} \label{oz}
h_{\text{ad}}(\rv_1,\rv_2, t) = 
c^{(2)}_{\text{ad}}(\rv_1,\rv_2, t) \\ + 
\int\! d\rv_3\, h_{\text{ad}}(\rv_1,\rv_3, t)\, 
\rho(\rv_3, t) \, 
c^{(2)}_{\text{ad}}(\rv_3,\rv_2, t),
\end{multline}
yields the adiabatic total correlation function 
$h_{\text{ad}}$, from which the adiabatic two-body density 
can easily be found using 
\begin{equation} \label{rho2 definition}
\rho^{(2)}_{\text{ad}}(\rv_1, \rv_2, t)
=
\rho(\rv_1, t) \rho(\rv_2, t)\big( 
\,h_{\text{ad}}(\rv_1, \rv_2, t)+1 \,\big).
\end{equation} 
We refer the reader to Ref. \cite{tschopp2} for details 
of how to implement the above scheme in cases for which the 
one-body density has either planar or spherical symmetry.


%

\end{document}